\documentclass{article}
\usepackage{waspaa23,amsmath,graphicx,url,times}

\usepackage{color}
\usepackage{multirow}
\usepackage{lipsum}

\usepackage{pifont}% http://ctan.org/pkg/pifont
\newcommand{\cmark}{\ding{51}}%
\newcommand{\xmark}{\ding{55}}%

\newcommand\blfootnote[1]{%
  \begingroup
  \renewcommand\thefootnote{}\footnote{#1}%
  \addtocounter{footnote}{-1}%
  \endgroup
}

\title{An objective evaluation of Hearing Aids and DNN-based binaural speech enhancement in complex acoustic scenes
}

\name{Enric Gus\'o,$^{1,2}$
      Joanna Luberadzka,$^{2}$
      Mart\'i Baig,$^{3}$
      Umut Sayin,$^{2}$
      Xavier Serra$^{1}$}
 \address{$^1$ Universitat Pompeu Fabra, Music Technology Group, Barcelona \\ enric.guso@upf.edu, xavier.serra@upf.edu\\
          $^2$ Eurecat, Centre Tecnològic de Catalunya, Tecnologies Multimèdia, Barcelona \\
           joanna.luberadzka@eurecat.org, umut.sayin@eurecat.org\\
          $^3$ Microson, Amplifon Group, Barcelona, marti.baig@amplifon.com\\
 }

\begin{document}

\ninept
\maketitle

\begin{sloppy}

\begin{abstract}
We investigate the objective performance of five high-end commercially available Hearing Aid (HA) devices compared to DNN-based speech enhancement algorithms in complex acoustic environments. To this end, we measure the HRTFs of a single HA device to synthesize a binaural dataset for training two state-of-the-art causal and non-causal DNN enhancement models. We then generate an evaluation set of realistic speech-in-noise situations using an Ambisonics loudspeaker setup and record with a KU100 dummy head wearing each of the HA devices, both with and without the conventional HA algorithms, applying the DNN enhancers to the latter. We find that the DNN-based enhancement outperforms the HA algorithms in terms of noise suppression and objective intelligibility metrics.
\end{abstract}


\section{Introduction}
\label{sec:intro}

\blfootnote{The research leading to these results has received funding from the European union's Horizon Europe programme under grant agreement No 101017884 - GuestXR project.}

Hearing Aids are electronic devices that attempt to compensate for hearing loss by means of frequency-dependent amplification and dynamic range compression among other techniques. Traditionally, in order to improve the signal-to-noise (SNR) ratio, HAs make use of directional microphones \cite{ricketts1999comparison} or adaptive beamformers \cite{gode2022adaptive}, usually evaluated with stationary noises under laboratory conditions \cite{IEC6011816}. However, real-life scenarios can be much more challenging due to the possible presence of non-stationary noises, interfering speech and/or several competing talkers. %Attempts have been made to simulate a more realistic acoustic environment in the laboratory for hearing aid testing in search of the ecological validity \cite{minnaar2013vse} ...}, 

Beyond the research in HA field, end-to-end deep neural networks (DNNs) have proven to be very successful in such challenging scenarios, even without exploiting spatial cues and focusing on the monaural case \cite{reddy2021icassp}, but these models are not directly suitable for HA use due to being non-causal (i.e. requiring future samples) or needing lots of computational power. Therefore, causal adaptations have been proposed in \cite{tzinis2020sudo} and further explored in \cite{ali2023scaling}, while in \cite{han2020real} Han et al. have adapted them to the binaural case while preserving spatial cues. In addition, \cite{wang2023neural} have achieved very low algorithmic latency that is comparable to HA processing \cite{stone2008delay} which, together with advances in the miniaturization of DNNs \cite{banbury2021micronets}, methods that offload the processing to other devices \cite{sun2020supervised}, and the outcomes of the Clarity Challenge \cite{clarity} constitute major steps towards more powerful DNN-based HA processing.

In fact, some HA manufacturers already use DNN processing as part of a post-filtering step in their products \cite{oticon1, oticon3, oticon2}, although they do not disclose any details of the actual architecture and implementation, nor ablation studies that quantify the performance gain from the DNNs. This work focuses precisely on evaluating this gap by using an experimental setup that is controlled but also as realistic and ecologically-valid as possible. Our approach has been to record several HAs
in realistic noisy environments and to compare their performance
with the same recordings post-processed by the DNNs. We have chosen \textit{Sudo-RM-RF} \cite{tzinis2020sudo}, an already developed, well-tested state of the art DNN originally designed for speech separation that has also shown good performance in the HA speech enhancement case \cite{leecitear, clarity}. The main contribution of this study is an objective evaluation which suggests that hearing aid signal enhancement has difficulties under challenging acoustic conditions and that DNN-based speech enhancement methods have the potential of improving current hearing devices. Furthermore, we provide the following research materials:
\begin{itemize}
    \item An Ambisonics to binaural decoder, built from a set of measured HRTFs from the KU100 wearing an \textit{audifon lewi R} ---a commercially-available HA Receiver-In-Canal (RIC) device.
    \item A new, synthetic and binaural (two-channel) HA speech enhancement training dataset built with that decoder.
    \item A reverberant speech-in-noise test set in Ambisonics and binaural formats, with the corresponding recordings for each HA device.
\end{itemize}

\section{HA MEASUREMENT SETUP}
\label{sec:ha}

Our measurement setup (depicted in Figure \ref{fig:results}) was designed to capture the signals processed by various HA devices within complex acoustic scenes. We recorded with five high-end RIC HAs available in the market at the time of writing: GN ONE 961-DRWC, GN ONE 561-DRWC, Phonak Audéo P90-R, Phonak Audéo P70-R, and Signia Pure C$\&$G 3x. In total, we have tested fifteen combinations of HA and receiver, recording with low, mid and high power receivers for each device.

\begin{figure}[t]
  \centering
  \centerline{\includegraphics[width=\columnwidth]{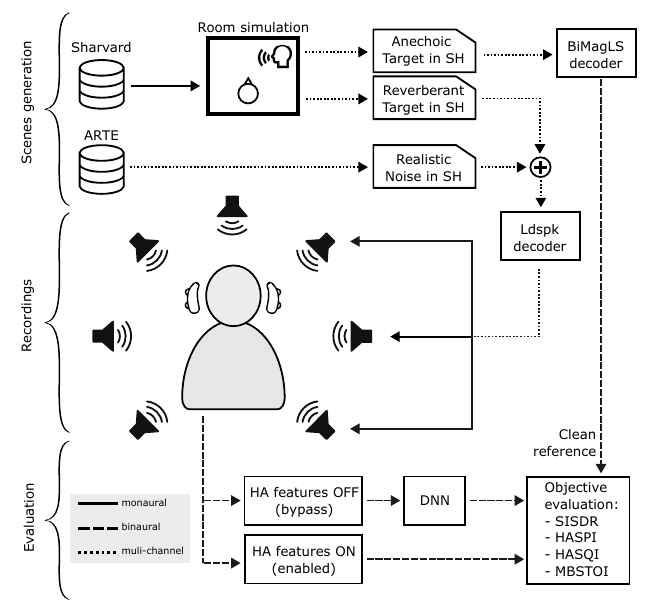}}
  \caption{Hearing aid measurement setup. Scenes generation: complex acoustic scenes are generated by combining existing databases (Sharvard, ARTE) with room acoustic simulation. Recordings: audio material is played back in an Ambisonics-based reproduction system and the signals processed by the HA are captured with the microphones of a dummy head. HA are recorded with and without signal-enhancing features. Evaluation: HA-enhanced recordings are compared with DNN-processed recordings using a range of objective metrics.}
  \label{fig:results}
\end{figure}

\textbf{Scenes generation ---} To ensure that the sound processed by the tested hearing devices closely resembled real-life scenarios, we used an Ambisonics-based spatial sound reproduction system. All HA devices were exposed to three different acoustic scenes of speech in noise. To create these scenes we used three noise recordings from the ARTE database \cite{weisser2019ambisonic} representing common sound environments: \textit{party}, \textit{restaurant} and \textit{office}. Recordings come from real environments captured with a 62-channel microphone array, and are available as 31-channel mixed-order Ambisonics signals  which we zero-padded up to 10th order. The target speech consisted of nine randomly selected sentences spoken by a female speaker from the Sharvard database \cite{aubanel2014shavrvard}. To simulate room acoustics, we used an adaptation of the Multichannel Acoustic Signal Processing Library\footnote{https://github.com/andresperezlopez/masp} (MASP): a shoebox room impulse response simulator based on the Image Source Method that allows for Spherical Harmonics expansion (SH, i.e. Ambisonics). We used 10th-order Ambisonics, which provide sufficient spatial resolution and generated two sets of sound fields at the left ear, right ear, and $head$ positions: one set where we only simulated sound propagation (the direct sound field w/o reflections) and the other one being the actual reverberation. We used 17.5cm of ear distance for computing the left and right ear coordinates from the $head$ origin to match the KU100 ear distance. We simulated three rooms with dimensions set to 15x10x3.5m, 28x17x4.2m, and 5x2x2.5m for the \textit{party}, \textit{restaurant}, and \textit{office} environments respectively. All targets where placed at 1m from the $head$ with two different angles: 0 degrees or 30 degrees to the right (relative to the head horizontal orientation $head_{\theta}$).  We adjusted the RT60 parameter using informal listening by comparing with the ARTE recordings. We chose RT60s that were $60\%$ of the ones reported in ARTE to account for furniture and people absorption. 

\textbf{Decoding ---} Despite in \cite{oldenburg} Thiemann et al. published HRTFs of the microphones of a BTE HA placed at a head and torso simulator, no recordings were made with KU100 dummy which was available for our evaluation. Besides, in this study, we had no access to the microphone signals due to evaluating commercially-available HA. Hence, we decided to measure our own set of HRTFs in a setup as close as possible to the intended HA recording setup (i.e. in the same room, with a HA coupled to the dummy head ear canal, w/o signal enhancement features and only providing a linear gain). We used a pair of \textit{audifon lewi R} HA devices and the HRTF sets were measured using the sweep method with a single Genelec 8020 loudspeaker following a 50-point Lebedev grid. Impulse responses were cropped before the arrival of the first wall reflection and low frequencies were extended by LFE algorithm \cite{lfe}. This set of HRTFs was then used to build the 10-th order Ambisonics to binaural decoder following the Bilateral Magnitude Least Squares method (BiMagLS) \cite{bimagls}, applying high order tapering with a cutoff frequency of 6239Hz ---which is the theoretical cutoff frequency for correct representation in 10-th order Ambisonics. To obtain the clean anechoic reference signal needed for the objective evaluation we took the left and right ear anechoic sound fields simulated in MASP and applied the BiMagLS decoder. We weighted the Ambisonics signals so that SNR was +5dB when decoded to binaural. Finally, we added the weighted speech and noise sound fields simulated at the $head$ center position, and decoded the resulting Ambisonics mixture into the loudspeaker signals using a fifth order in-phase decoder optimized with IDHOA \cite{scaini2014decoding}, particularly tailored to our specific loudspeaker setup. We also normalized all sets of speaker signals to have the same energy (sum of squares) for ease of calibration.

\textbf{Recordings ---} The KU100 dummy head was positioned at the center of a three-dimensional irregular loudspeaker array comprising 25 Genelec 8040s loudspeakers. We calibrated the system so all scenes were at 70dB SPL. For each recording, we placed the hearing aids behind the ears of the KU100 dummy head and inserted the receiver into the ear canal. To minimize the influence of direct sound, we occluded the entrance to the ear canal with adhesive putty material in addition to the HA's power dome. 

We recorded each hearing device in two modes: \textit{bypass} and \textit{enabled}. In \textit{bypass} mode all the HA algorithms were deactivated except for the feedback canceller and a linear amplification of approximately 20dB. Hearing aid models chosen for this study are commercially available HA. For such devices it is not straightforward to record directly from the HA microphone. Instead, we used signals recorded in \textit{bypass} mode at the KU100 as an approximation of the HA microphone signals. The \textit{bypass} recordings were used as the input to the offline DNN-based speech enhancement methods. In contrast to the \textit{bypass} recordings, in the \textit{enabled} mode the HA applied the signal enhancement algorithms present in the default factory settings. In all tested devices, these settings included at least some form of adaptive beamforming and single-channel noise reduction. No hearing correction was applied. 

In a preliminary round of recordings we employed the phase-inversion procedure \cite{hagerman2004method} commonly used to estimate the SNR at the output of a linear hearing aid. However, we noticed that for some of the HA and for all DNN-based algorithms the linearity assumption of the method could not be met, making the SNR estimates obtained with this method unreliable. Therefore, we decided to rely on intrusive, reference-based metrics instead. 

% Please add the following required packages to your document preamble:
% \usepackage{multirow}
\begin{table}
\centering
\begin{tabular}{|c|cc|cc|ccc|}
\hline
\multirow{2}{*}{set} & \multicolumn{2}{c|}{MLSS}                                            & \multicolumn{2}{c|}{WHAM!}      & \multicolumn{3}{c|}{noise  augmentations}                           \\
                     & \#                                         & hours                  & \#                       & hours & $\Phi inv$                  & L$\Leftrightarrow$R         & stretch \\ \hline
\multirow{6}{*}{\textit{tr}}  & \multicolumn{1}{c|}{\multirow{6}{*}{221k}} & \multirow{6}{*}{245.6} & \multicolumn{1}{c|}{52k} & 57.8 & \multicolumn{1}{c|}{\xmark} & \multicolumn{1}{c|}{\xmark} & \xmark  \\ \cline{4-8} 
                     & \multicolumn{1}{c|}{}                      &                        & \multicolumn{1}{c|}{52k} & 57.8 & \multicolumn{1}{c|}{\cmark} & \multicolumn{1}{c|}{\xmark} & \xmark  \\ \cline{4-8} 
                     & \multicolumn{1}{c|}{}                      &                        & \multicolumn{1}{c|}{52k} & 57.8 & \multicolumn{1}{c|}{\xmark} & \multicolumn{1}{c|}{\cmark} & \xmark  \\ \cline{4-8} 
                     & \multicolumn{1}{c|}{}                      &                        & \multicolumn{1}{c|}{52k} & 57.8 & \multicolumn{1}{c|}{\cmark} & \multicolumn{1}{c|}{\cmark} & \xmark  \\ \cline{4-8} 
                     & \multicolumn{1}{c|}{}                      &                        & \multicolumn{1}{c|}{6.5k} & 7.2  & \multicolumn{1}{c|}{\xmark} & \multicolumn{1}{c|}{\xmark} & \cmark  \\ \cline{4-8} 
                     & \multicolumn{1}{c|}{}                      &                        & \multicolumn{1}{c|}{6.5k} & 7.2  & \multicolumn{1}{c|}{\cmark} & \multicolumn{1}{c|}{\cmark} & \cmark  \\ \hline
\textit{cv}                   & \multicolumn{1}{c|}{2.4k}                   & 2.67                   & \multicolumn{1}{c|}{2.4k} & 2.67 & \multicolumn{1}{c|}{\xmark} & \multicolumn{1}{c|}{\xmark} & \xmark  \\ \hline
\textit{tt}                   & \multicolumn{1}{c|}{2.4k}                   & 2.67                   & \multicolumn{1}{c|}{2.4k} & 2.67 & \multicolumn{1}{c|}{\xmark} & \multicolumn{1}{c|}{\xmark} & \xmark  \\ \hline
\end{tabular}
\caption{Training data splits and augmentations, where \textit{tr} stands for the training set, \textit{cv} for validation, \textit{tt} for the test set, \# for the number of utterances, $\Phi inv$ for changing the sign of the noise signal, L$\Leftrightarrow$R for permuting the noise channels, and \textit{stretch} for applying time stretching with a random factor.}
\label{tab:dataset}
\end{table}

\textbf{Evaluation ---} Objective evaluation compared the conventional HA enhancement algorithms with the DNNs. We evaluated four sets of recordings: \textit{bypass}, \textit{enabled}, \textit{bypass} post-processed with DNN and \textit{bypass} post-processed with DNN-C. The first set represents recordings without signal enhancement and the remaining three sets represent the different signal enhancement strategies. For each set we computed four objective metrics: Hearing-Aid Speech Quality Index (HASQI) \cite{hasqi}, Hearing-Aid Speech Perception Index (HASPI) \cite{haspi}, Modified Binaural Short-Time Objective Intelligibility (MBSTOI) \cite{andersen2018refinement} and Scale-invariant signal-to-distortion ratio (SISDR) as in \cite{tzinis2020sudo}. Given the clean anechoic target binaural speech signal $y$, the DNN or HA estimate $\tilde{y}$ and a baseline recording with the HA in \textit{bypass} $\hat{y}$ (all $\tau$ samples long) SISDR is described in Equation \ref{sisdr}. $y$ was used as common reference for all metrics. Reference and estimate pairs were time-aligned using the cross-correlation method. We took the best ear for the non-binaural measures (SISDR, HASPI, HASQI) and normalized the signals as in \cite{clarity}. We used a flat normal-hearing audiogram as an input to HASPI and HASQI metrics. We define the signal enhancement benefit as the difference in objective metrics between the non-enhanced and enhanced signals. For example, $\Delta \text{SISDR} = \text{SISDR}(\tilde{y}_{t}, y_{t}) - \text{SISDR}(\hat{y}_{t}, y_{t})$. The rest of metrics expressing the signal enhancement benefit are denoted as $\Delta$HASQI, $\Delta$HASPI, $\Delta$MBSTOI and computed in the same fashion.

\begin{equation}
\label{sisdr}
{\text{SISDR}(\tilde {y}_{t}, y_{t})}=\frac{10}{\scalebox{1.5}{$\tau$}} \sum_{t} \log_{10}\left(\frac{\left|\frac{\tilde{y}_{t}^{\text{T}}y_{t}}{|y_{t}|^{2}}y_{t}\right|^{2}}{\left|\frac{\tilde{y}_{t}^{T}y_{t}}{|y_{t}|^{2}}y_{t}-\tilde{y}_{t}\right|^{2}}\right) 
\end{equation}

\section{DNN DATASET AND TRAINING SETUP}
\label{sec:dnn_model}

\textbf{Task ---} We approach supervised binaural speech enhancement with DNNs, which requires a large number of noisy and reverberant mixtures, as well as the corresponding targets (clean speech) also in binaural to preserve spatial cues. Both inputs and targets have to resemble the ones that would be recorded with the two frontal omnidirectional microphones in a Behind The Ear (BTE) HA. To accomplish this, we simulate reverberation in the Ambisonics spatial audio domain and then decode to binaural signals by using a decoder specifically-tailored for HA.

\textbf{Datasets ---} As in \cite{reddy2021icassp, tzinis2020sudo}, we relied on speech recordings from audiobooks, in this case using the Spanish subset from Multilingual LibriSpeech (MLSS) dataset \cite{pratap2020mls} for the clean signals. We used a sampling rate of 16kHz and took four seconds chunks, selecting the chunk that presented more energy in order to avoid silence, obtaining $22 \cdot 10^{5}$ utterances for the training set, 2408 for validation and 2385 for testing, which add up to 251 hours of clean speech. Regarding noise, we used the WHAM! \cite{wichern2019wham} binaural dataset which contains babble speech, cafeteria noise and background music. We kept the original data splits from both datasets, preserving gender balance and avoiding contamination between sets. We augmented WHAM! training set to match the length of MLSS by following three main strategies: \textit{{i)}} we flipped the phase, \textit{{ii)}} we swapped left and right channels as a rough approximation of a 180º rotation in the horizontal plane and \textit{{iii)}} we randomly time-stretched from 90\% to 110\% of the noise duration. For validation and testing we randomly picked from WHAM! validation and test splits respectively. Details on the data splits is shown in Table \ref{tab:dataset}.

\textbf{Room simulation ---} MLSS contains close-mic studio recordings, so we have considered them to be a fair approximation of anechoic signals. The random room configuration details are shown in Table \ref{tab:rooms}. We used MASP and our Ambisonics to binaural HA decoder as in Section \ref{sec:ha}. In an attempt to make our models agnostic to our particular HA and dummy combination, the response of the RIC coupling was compensated by taking the direction-independent frequency response between KU100 and KU100 wearing the HA HRTFs and approximating it with an IIR filter that was applied to all utterances in the dataset.

\begin{table}
\centering
\begin{tabular}{l|l}
\hline
                                             &                                                     \\
$r_{x}=\mathcal{U}(3 ,  30)$                 & $head_x=\mathcal{U}(0.35r_{x}, 0.65r_{x})$          \\
$r_{y}=r_{x} \cdot \mathcal{U}(0.5 ,  1) $   & $head_y=\mathcal{U}(0.35r_{y}, 0.65r_{y})$          \\
$r_{z}=\mathcal{U}(2.5,  5)   $              & $head_z=\mathcal{U}(1, 2)$                          \\
                                             &                                                     \\
$||head - target|| =\mathcal{U}(0.5, 3)$     & $head_{\theta}=\mathcal{U}(-45, 45)$                \\
$\angle head, target = \mathcal{U}(-45, 45)$ & $head_{\psi}=\mathcal{U}(-10, 10)$                  \\
$\text{SNR}=\mathcal{U}(0,6)$                       & $\text{RT60} =\mathcal{U}(0.1, 0.5) \frac{\text{SNR} + 0.3}{5.3}   $ \\
                                             &                                                     \\ \hline
\end{tabular}
\caption{Random room configuration, sampled from uniform distribution $\mathcal{U} $. Room $r$ and $head$ dimensions and distances are in meters, SNR in dBs, RT60 in seconds and $head$ azimuth $\theta$, elevation $\psi$ and angle with target in degrees. $r_{y}$ depends on $r_{x}$ to avoid corridor-like spaces, and RT60 depends on the SNR because the more noisy a situation, the less reverberant it should be due to the crowd's absorption.}
\label{tab:rooms}
\end{table}
\textbf{Training Setup ---} Regarding the DNN topology, we did not make any modifications to \textit{SuDo-RM-RF} on top of configuring its encoder for receiving stereo audio. All network topology details can be found in \cite{tzinis2020sudo}. We trained two different models: DNN (which corresponds to \textit{SuDoRM-RF++GC} in their paper), a non-causal improved version that serves us as upper baseline, and DNN-C (the causal, HA oriented version that corresponds to \textit{C-SuDoRM-RF++} in their paper). All hyperparameters are shared between DNN and DNN-C except for the batch size ---which had to be reduced from 12 for DNN-C to 2 for DNN to fit into VRAM. We used 256 input and 512 output channels on 16 successive blocks with five upsampling/downsampling layers each, an encoder and decoder kernel size of 21 generating embeddings with a length of 512, four attention heads with 256 depth and 0.1 dropout (applied only during training) and an Adam optimizer. Learning rate was $10^{-3}$ and was divided by 3 every 8 epochs. We trained the whole dataset for 25 epochs. New mixtures and targets were generated every epoch by permuting the reverberant speech and the noise in every batch, and were normalized to zero mean and unit variance. We used SISDR as loss function and also as evaluation metric and obtained a test set performance of 11.7dB.

\section{RESULTS}
\label{sec:PDF_express}

Figure \ref{fig:results1} displays SISDR, HASPI, HASQI, and MBSTOI metrics obtained for each hearing device averaged across receiver types. In the \textit{bypass} condition metrics range from -12.63dB to -8.01dB for SISDR, 0.57 to 0.73 for HASPI, 0.14 to 0.16 for HASQI, and 0.39 to 0.49 for MBSTOI. 
Enabling signal enhancement features in hearing aids can have a different effect on their performance depending on the specific device. For example, device 4 demonstrates improvement of 0.73dB on SISDR, 0.14 on HASPI 0.02 on HASQI and 0.01 on MBSTOI while device 2 presents -0.54dB SISDR, -0.17 HASPI, -0.07 HASQI and -0.05 MBSTOI. Apart from these slight deviations, the overall impact of HA features is typically insignificant. In contrast, significant change can be observed in recordings processed with the DNNs. For the non-causal DNN, values reach -6.89dB to -1.21dB for SISDR, 0.60 to 0.77 HASPI, 0.14 to 0.21 for HASQI  and 0.57 to 0.69 for MBSTOI.

Figure \ref{fig:results2} depicts the signal enhancement benefit. Violin plots summarize the values obtained from the 15 hearing devices averaged along all scenes. For hearing aid features the benefit is on average 0.014 dB for SISDR, -0.02 for HASQI, -0.01 for HASPI, and -0.01 for MBSTOI. A much larger improvement could be observed for DNN-based enhancement. For a causal model, the mean benefit was 4.96 dB for SISDR, -0.01 for HASQI, 0.02 for HASPI, and 0.15 for MBSTOI. The largest benefit was achieved by the non-causal DNN model, with mean values reaching 6.09 dB for SISDR, 0.02 for HASQI, 0.07 for HASPI, and 0.17 for MBSTOI.

\begin{figure}
  \centering
  \centerline{\includegraphics[width=\columnwidth, trim={0.5cm 1cm 0.5cm 1cm}]{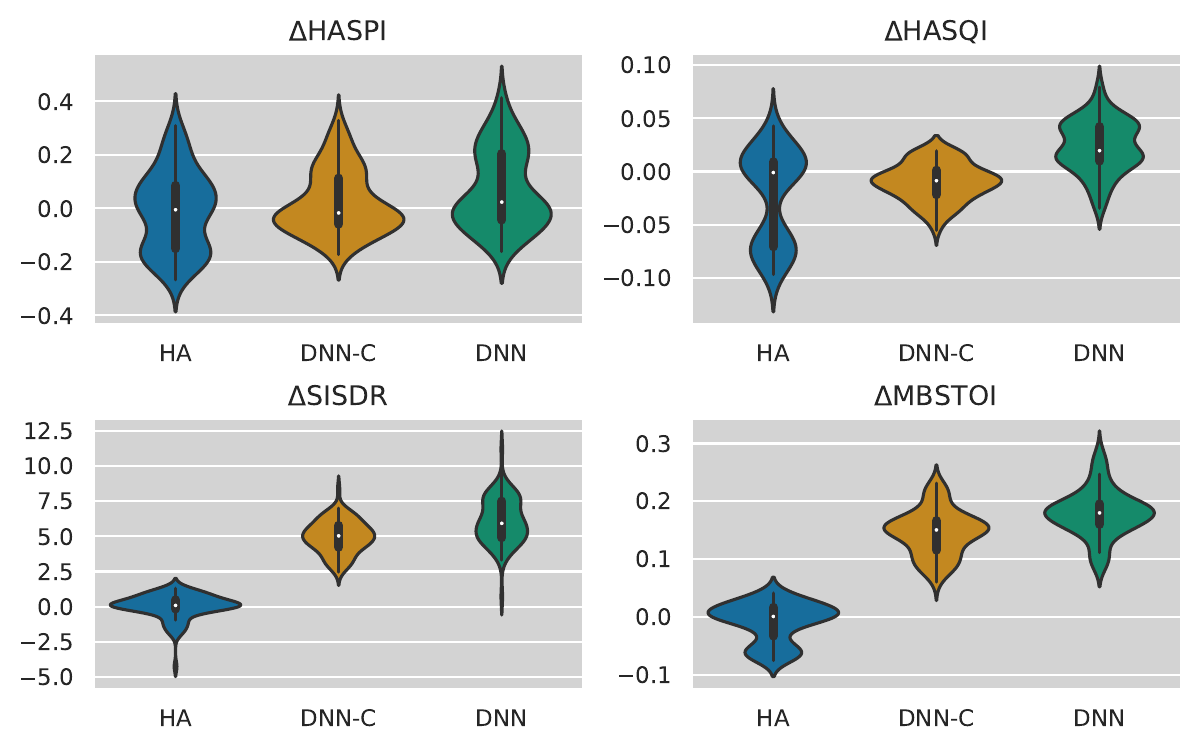}}
  \caption{Signal enhancement benefit for hearing aids (HA), non-causal DNN model (DNN) and causal DNN model (DNN-C).}
  \label{fig:results2}
\end{figure}

\begin{figure}[h!]
  \centering
  \centerline{\includegraphics[width=0.8\columnwidth, trim={0 1cm 0 1cm}]{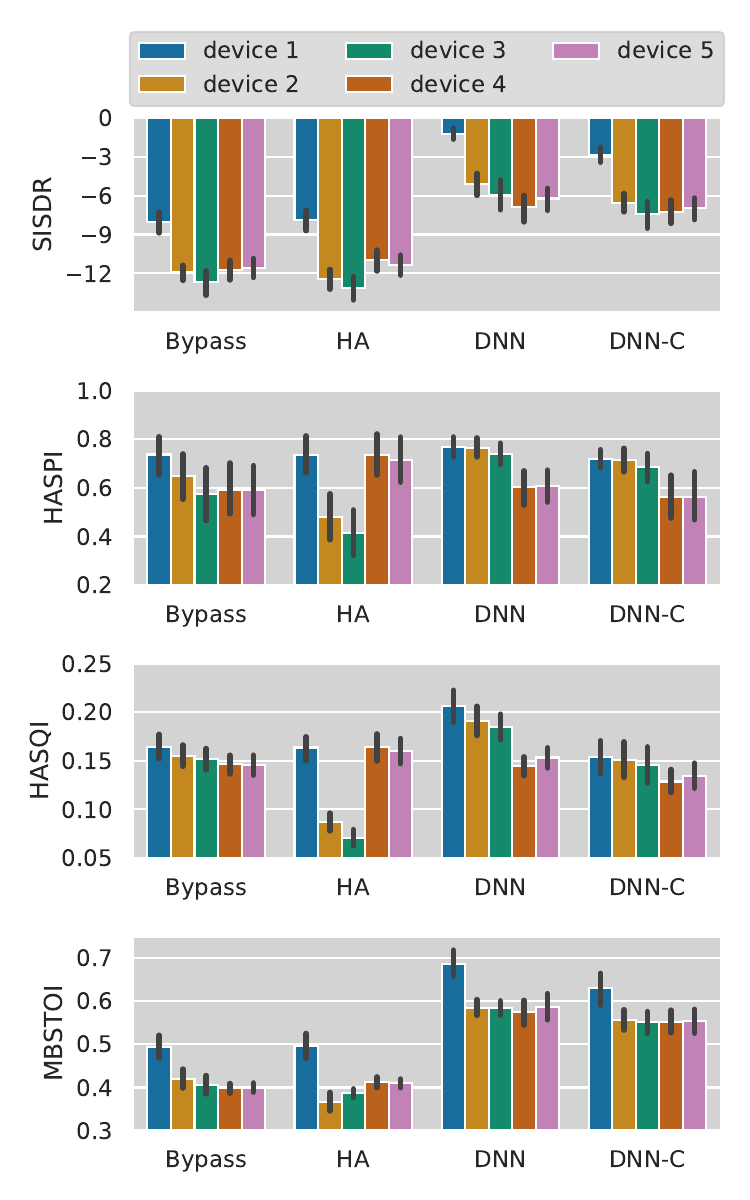}}
  \caption{Objective metrics across devices, anonymized for avoiding any drawal of inappropriate interdevice conclusions.}
  \label{fig:results1}
\end{figure}

\section{DISCUSSION}
In this paper we have presented a setup which allows to quantify the performance gap between the current hearing aid signal enhancement strategies and DNN-based approaches. On the one hand, HA processing seems to struggle in these complex situations, even performing worse than in \textit{bypass} for some devices, perhaps because beamformers cannot estimate the direction of the target speech due to the noise and competing talkers being non-stationary. On the other hand, this does not seem to affect DNNs, which improve SISDR and MBSTOI consistently. 

Interestingly, differences between devices in \textit{bypass} are still observed after being processed by the DNNs, suggesting that DNN models are sensitive to the quality of the inputs. It would be interesting to study the concatenation of DNN and traditional strategies in future work.

We acknowledge that using the binaural decoding of the anechoic signal as reference can seem counter-intuitive because HA are not designed to provide anechoic estimates yet. However, this reference was the one that matched better with our informal listening. Another limitation of the present study is that we are comparing real time HA processing with DNNs applied as a post-processing, because we found it difficult to obtain signal insert points in consumer HA devices and also because no real time implementations of these models are publicly available yet. However, the slight performance difference between DNN and DNN-C and both being far better than the HA in terms of SISDR and MBSTOI should additionally encourage this research direction. 

\section{conclusions}
We have shown that HA enhancement algorithms struggle in eco- logically valid complex situations reproduced at the studio, even to the point of damaging performance compared to not applying any algorithm at all. We have also shown that DNN-based approaches have the potential of outperforming them in terms of denoising and intelligibility at the expense of quality, encouraging future work on optimization and pruning this kind of algorithms.
%We suggest to evaluate this subjectively with hearing impaired in future work. 

%\item "No transparency of hearing devices" + "Problems for putting devices in bypass" 
%We did not record the microphone signals from the hearing aid but the hearing aid in bypass which should be just a linearly amplified signal but in reality for some devices (which we excluded from the paper), it was actually doing some adaptive processing. We believe the remaining recordings are a good approximation of the hearing aid mic but in the future, there is a need for a more transparent hearing devices, so in the future we could use the stuff from Oldenburg (portable hearing lab)...

%\item should be nice to also try easy scenarios (for which the HA has been designed) with DNN
%\item one SNR... 

%\subsection{Subheadings}
%\label{ssec:subhead}

%\section{Equations}
%\label{sec:equations}

% -------------------------------------------------------------------------
% Either list references using the bibliography style file IEEEtran.bst

\clearpage
\newpage

\bibliographystyle{IEEEtran}
\bibliography{refs23}

\begin{thebibliography}{10}
\providecommand{\url}[1]{#1}
\def\UrlFont{\rmfamily}
\providecommand{\newblock}{\relax}
\providecommand{\bibinfo}[2]{#2}
\providecommand\BIBentrySTDinterwordspacing{\spaceskip=0pt\relax}
\providecommand\BIBentryALTinterwordstretchfactor{4}
\providecommand\BIBentryALTinterwordspacing{\spaceskip=\fontdimen2\font plus
\BIBentryALTinterwordstretchfactor\fontdimen3\font minus
  \fontdimen4\font\relax}
\providecommand\BIBforeignlanguage[2]{{%
\expandafter\ifx\csname l@#1\endcsname\relax
\typeout{** WARNING: IEEEtran.bst: No hyphenation pattern has been}%
\typeout{** loaded for the language `#1'. Using the pattern for}%
\typeout{** the default language instead.}%
\else
\language=\csname l@#1\endcsname
\fi
#2}}

\bibitem{ricketts1999comparison}
T.~Ricketts and S.~Dhar, ``Comparison of performance across three directional
  hearing aids,'' \emph{Journal of the American Academy of Audiology}, vol.~10,
  no.~04, pp. 180--189, 1999.

\bibitem{gode2022adaptive}
H.~Gode and S.~Doclo, ``Adaptive dereverberation, noise and interferer
  reduction using sparse weighted linearly constrained minimum power
  beamforming,'' in \emph{2022 30th European Signal Processing Conference
  (EUSIPCO)}.\hskip 1em plus 0.5em minus 0.4em\relax IEEE, 2022, pp. 95--99.

\bibitem{IEC6011816}
``Electroacoustics –- hearing aids –- part 16: Definition and verification
  of hearing aid features,'' International Electrotechnical Commission, Geneva,
  CH, Standard, Mar. 2022.

\bibitem{reddy2021icassp}
C.~K. Reddy, H.~Dubey, V.~Gopal, R.~Cutler, S.~Braun, H.~Gamper, R.~Aichner,
  and S.~Srinivasan, ``Icassp 2021 deep noise suppression challenge,'' in
  \emph{ICASSP 2021-2021 IEEE International Conference on Acoustics, Speech and
  Signal Processing (ICASSP)}.\hskip 1em plus 0.5em minus 0.4em\relax IEEE,
  2021, pp. 6623--6627.

\bibitem{tzinis2020sudo}
E.~Tzinis, Z.~Wang, X.~Jiang, and P.~Smaragdis, ``Compute and memory efficient
  universal sound source separation,'' \emph{Journal of Signal Processing
  Systems}, vol.~94, no.~2, pp. 245--259, 2022.

\bibitem{ali2023scaling}
M.~N. Ali, F.~Paissan, D.~Falavigna, and A.~Brutti, ``Scaling strategies for
  on-device low-complexity source separation with conv-tasnet,'' \emph{arXiv
  preprint arXiv:2303.03005}, 2023.

\bibitem{han2020real}
C.~Han, Y.~Luo, and N.~Mesgarani, ``Real-time binaural speech separation with
  preserved spatial cues,'' in \emph{ICASSP 2020-2020 IEEE International
  Conference on Acoustics, Speech and Signal Processing (ICASSP)}.\hskip 1em
  plus 0.5em minus 0.4em\relax IEEE, 2020, pp. 6404--6408.

\bibitem{wang2023neural}
Z.-Q. Wang, S.~Cornell, S.~Choi, Y.~Lee, B.-Y. Kim, and S.~Watanabe, ``Neural
  speech enhancement with very low algorithmic latency and complexity via
  integrated full- and sub-band modeling,'' 2023.

\bibitem{stone2008delay}
M.~A. Stone, B.~C. Moore, K.~Meisenbacher, and R.~P. Derleth, ``Tolerable
  hearing aid delays. v. estimation of limits for open canal fittings,''
  \emph{Ear and hearing}, vol.~29, no.~4, pp. 601--617, 2008.

\bibitem{banbury2021micronets}
C.~Banbury, C.~Zhou, I.~Fedorov, R.~Matas, U.~Thakker, D.~Gope,
  V.~Janapa~Reddi, M.~Mattina, and P.~Whatmough, ``Micronets: Neural network
  architectures for deploying tinyml applications on commodity
  microcontrollers,'' \emph{Proceedings of machine learning and systems},
  vol.~3, pp. 517--532, 2021.

\bibitem{sun2020supervised}
Z.~Sun, Y.~Li, H.~Jiang, F.~Chen, X.~Xie, and Z.~Wang, ``A supervised speech
  enhancement method for smartphone-based binaural hearing aids,'' \emph{IEEE
  Transactions on Biomedical Circuits and Systems}, vol.~14, no.~5, pp.
  951--960, 2020.

\bibitem{clarity}
M.~A. Akeroyd, W.~Bailey, J.~Barker, T.~J. Cox, J.~F. Culling, S.~Graetzer,
  G.~Naylor, Z.~Podwi{\'n}ska, and Z.~Tu, ``The 2nd clarity enhancement
  challenge for hearing aid speech intelligibility enhancement: Overview and
  outcomes,'' in \emph{ICASSP 2023-2023 IEEE International Conference on
  Acoustics, Speech and Signal Processing (ICASSP)}.\hskip 1em plus 0.5em minus
  0.4em\relax IEEE, 2023, pp. 1--5.

\bibitem{oticon1}
A.~H. Andersen, S.~Santurette, M.~S. Pedersen, E.~Alickovic, L.~Fiedler,
  J.~Jensen, and T.~Behrens, ``Creating clarity in noisy environments by using
  deep learning in hearing aids,'' in \emph{Seminars in Hearing}, vol.~42,
  no.~03.\hskip 1em plus 0.5em minus 0.4em\relax Thieme Medical Publishers,
  Inc., 2021, pp. 260--281.

\bibitem{oticon3}
M.~Brændgaard, ``Tech paper: An introduction to moresound intelligence,''
  Centre for Applied Audiology Research, Oticon A/S, Tech. Rep., 2020.

\bibitem{oticon2}
S.~Santurette, L.~Xia, C.~A. Ermert, and B.~Man Kai~Loong, ``Whitepaper: Oticon
  more competitive benchmark - part 1 – technical evidence,'' Centre for
  Applied Audiology Research, Oticon A/S, Tech. Rep., 2021.

\bibitem{leecitear}
C.-C. Lee, H.-W. Chen, R.~Chao, T.-T. Liu, and Y.~Tsao, ``Citear: A two-stage
  end-to-end system for noisy-reverberant hearing-aid processing,'' in
  \emph{Proc. Clarity-CEC2-2022}, 2022.

\bibitem{weisser2019ambisonic}
A.~Weisser, J.~M. Buchholz, C.~Oreinos, J.~Badajoz-Davila, J.~Galloway,
  T.~Beechey, and G.~Keidser, ``The ambisonic recordings of typical
  environments (arte) database,'' \emph{Acta Acustica United With Acustica},
  vol. 105, no.~4, pp. 695--713, 2019.

\bibitem{aubanel2014shavrvard}
V.~Aubanel, M.~L.~G. Lecumberri, and M.~Cooke, ``The sharvard corpus: A
  phonemically-balanced spanish sentence resource for audiology,''
  \emph{International journal of audiology}, vol.~53, no.~9, pp. 633--638,
  2014.

\bibitem{oldenburg}
J.~Thiemann and S.~van~de Par, ``A multiple model high-resolution head-related
  impulse response database for aided and unaided ears,'' \emph{EURASIP Journal
  on Advances in Signal Processing}, vol. 2019, pp. 1--9, 2019.

\bibitem{lfe}
B.~Bernsch{\"u}tz, ``A spherical far field hrir/hrtf compilation of the neumann
  ku 100,'' in \emph{Proceedings of the 40th Italian (AIA) annual conference on
  acoustics and the 39th German annual conference on acoustics (DAGA)
  conference on acoustics}.\hskip 1em plus 0.5em minus 0.4em\relax German
  Acoustical Society (DEGA) Berlin, 2013, p.~29.

\bibitem{bimagls}
I.~Engel, D.~Goodman, and L.~Picinali, ``Improving binaural rendering with
  bilateral ambisonics and magls,'' in \emph{Annual German Conference on
  Acoustics}, vol.~99, no.~1, 2021, p.~10.

\bibitem{scaini2014decoding}
D.~Scaini and D.~Arteaga, ``Decoding of higher order ambisonics to irregular
  periphonic loudspeaker arrays,'' in \emph{Audio Engineering Society
  Conference: 55th International Conference: Spatial Audio}.\hskip 1em plus
  0.5em minus 0.4em\relax Audio Engineering Society, 2014.

\bibitem{hagerman2004method}
B.~Hagerman and {\AA}.~Olofsson, ``A method to measure the effect of noise
  reduction algorithms using simultaneous speech and noise,'' \emph{Acta
  Acustica United with Acustica}, vol.~90, no.~2, pp. 356--361, 2004.

\bibitem{hasqi}
J.~M. Kates and K.~H. Arehart, ``The hearing-aid speech quality index (hasqi)
  version 2,'' \emph{Journal of the Audio Engineering Society}, vol.~62, no.~3,
  pp. 99--117, 2014.

\bibitem{haspi}
------, ``The hearing-aid speech perception index (haspi) version 2,''
  \emph{Speech Communication}, vol. 131, pp. 35--46, 2021.

\bibitem{andersen2018refinement}
A.~H. Andersen, J.~M. de~Haan, Z.-H. Tan, and J.~Jensen, ``Refinement and
  validation of the binaural short time objective intelligibility measure for
  spatially diverse conditions,'' \emph{Speech Communication}, vol. 102, pp.
  1--13, 2018.

\bibitem{pratap2020mls}
V.~Pratap, Q.~Xu, A.~Sriram, G.~Synnaeve, and R.~Collobert, ``{MLS: A
  Large-Scale Multilingual Dataset for Speech Research},'' in \emph{Proc.
  Interspeech 2020}, 2020, pp. 2757--2761.

\bibitem{wichern2019wham}
G.~Wichern, J.~Antognini, M.~Flynn, L.~R. Zhu, E.~McQuinn, D.~Crow, E.~Manilow,
  and J.~L. Roux, ``{WHAM!: Extending Speech Separation to Noisy
  Environments},'' in \emph{Proc. Interspeech 2019}, 2019, pp. 1368--1372.

\end{thebibliography}
%
% or list them by yourself
% \begin{thebibliography}{9}
%
% \bibitem{waspaaweb}
%   \url{http://www.waspaa.com}.
%
% \bibitem{IEEEPDFSpec}
%   {PDF} specification for {IEEE} {X}plore$^{\textregistered}$,
%   \url{http://www.ieee.org/portal/cms_docs/pubs/confstandards/pdfs/IEEE-PDF-SpecV401.pdf}.
%
% \bibitem{PDFOpenSourceTools}
%   Creating high resolution {PDF} files for book production with
%   open source tools,
%   \url{http://www.grassbook.org/neteler/highres_pdf.html}.
%
% \bibitem{eWilliams1999}
% E. Williams, \emph{Fourier Acoustics: Sound Radiation and Nearfield Acoustic
%   Holography}. London, UK: Academic Press, 1999.
%
% \bibitem{ieeecopyright}
%   \url{http://www.ieee.org/web/publications/rights/copyrightmain.html}.
%
% \bibitem{cJones2003}
% C. Jones, A. Smith, and E. Roberts, ``A sample paper in conference
%   proceedings,'' in \emph{Proc. IEEE ICASSP}, vol. II, 2003, pp. 803--806.
%
% \bibitem{aSmith2000}
% A. Smith, C. Jones, and E. Roberts, ``A sample paper in journals,''
%   \emph{IEEE Trans. Signal Process.}, vol. 62, pp. 291--294, Jan. 2000.
%
% \end{thebibliography}

\end{sloppy}
\end{document}